\documentclass{vgtc}                          

\usepackage{times}                     
\usepackage{mathptmx}                  

\vgtcinsertpkg

\title{Bridging Quantitative and Qualitative Methods for Visualization Research: A Data/Semantics Perspective in Light of Advanced AI}

\author{Daniel Weiskopf\thanks{e-mail: weiskopf@visus.uni-stuttgart.de}\\ %
        \scriptsize VISUS, University of Stuttgart}

\abstract{
This paper revisits the role of quantitative and qualitative methods in visualization research in the context of advancements in artificial intelligence (AI). The focus is on how we can bridge between the different methods in an integrated process of analyzing user study data. To this end, a process model of---potentially iterated---semantic enrichment and transformation of data is proposed. This joint perspective of data and semantics facilitates the integration of quantitative and qualitative methods. The model is motivated by examples of own prior work, especially in the area of eye tracking user studies and coding data-rich observations. Finally, there is a discussion of open issues and research opportunities in the interplay between AI, human analyst, and qualitative and quantitative methods for visualization research.
}

\keywords{Quantitative methods, qualitative methods, mixed methods, artificial intelligence, coding.}

\begin{document}

\firstsection{Introduction}

\maketitle
In this position statement, I address the issue of evaluation methodology for visualization research by revisiting the long-standing and ongoing discussion of quantitative and qualitative research methods. With the rapid advances in machine learning (ML) and artificial intelligence (AI), there is already a massive impact on how qualitative research can be conducted, in particular, by using large-language models (LLMs) for analyzing a wide range of text documents acquired in typical qualitative research. 

I take the current and future developments in AI as a motivation to rethink the \emph{roles} and \emph{interplay of qualitative and quantitative} methods in visualization research. In particular, there is an opportunity to \emph{bridge} between the two general approaches to address the many, still open issues related to evaluation, thus providing a new facet to the goal of ``BEyond time and errors: novel evaLuation methods for Information Visualization.''\footnote{First BELIV Workshop 2006 on ``BEyond time and errors: novel evaLuation methods for Information Visualization'' (\url{https://beliv-workshop.github.io/2006/}),
updated name ``evaluation and BEyond -- methodoLogIcal approaches for Visualization (BELIV)'' for the 2024  Workshop (\url{https://beliv-workshop.github.io/about.html})}

Let us first take a step back, getting a very coarse view on research methodology. One observation is that \emph{quantitative data} (e.g., in the sense of collected floating-point numbers or similar) is almost everywhere in visualization evaluation---for qualitative and quantitative methods. In fact, qualitative research often acquires massive amounts of data, e.g., in the form of long video or audio recordings in observational studies or interviews. This (obvious) observation is a first handle to bridge between quantitative and qualitative methods. However, there are also good reasons why the two are usually quite separated. 

In a very simplified, abstracted, and exaggerated view, qualitative methods tend to work from rich sets of observations to build new information and insight from there, i.e., the \emph{study leads to semantic information}. For example, observations from ethnographic or field studies may be analyzed by employing methods from grounded theory or qualitative content analysis to arrive at condensed meaning and understanding.
A key point is coding in particular, as well as semantic extraction, summarization, and reasoning of the study data in general. The analysis perspective includes exploratory aspects.

In contrast, quantitative methods tend to design an experiment to check some hypothesis. A typical example is a controlled laboratory study with null hypothesis significance testing (NHST) to compare two means of measured quantities of task performance (time or error). Therefore, much of the semantics or understanding comes beforehand and guides the design of the study, i.e., \emph{model and semantics inform the experiment}. Data analysis typically relies on statistical techniques, often with inferential statistics and the goal of confirmatory analysis.

As a disclaimer, it should be pointed out that the above characterizations are simplifications to make a point of the basic differences. Of course, there is a gray zone of techniques in between, particularly including approaches from mixed methods. However, the exaggerated characterization allows for the explanation of how bridging between different approaches may be possible.
The goal of bridging is to combine the best of two worlds: (1) ways to generate new insight and explanations from observations, as well as support for accounting for large design spaces and uncontrolled setups (advantages of many qualitative methods), (2) improved validity (in particular, internal validity), measurements of relative importance, and preciseness and detail of results (advantages of typical quantitative methods).

It is the right time to revisit the relationship between quantitative and qualitative methods because the rapid advances in AI make it practically feasible to have efficient, often largely, or even solely computer-based analysis methods that allow us to code qualitative data or extract and count patterns with unprecedented ease. Another reason is the ever-increasing availability of various sources of data from user studies, including physiological sensing like eye tracking or electroencephalography, audio and video recordings, or interaction logging. Finally, there is an increased need to better understand the relationship between humans and machine (and teams thereof), with a trend toward joint work between AI agents and users. 

A key point is that data can be \emph{enriched with additional semantics}, particularly during the analysis of user study data. For example, we can assign gaze data from eye tracking studies to semantic areas of interest to bring meaning to the distribution of attention. Traditionally, manual coding has been used for semantic enrichment. However, in light of advanced AI and the abundance of \emph{data-rich observations} from user studies, the important step of semantic enrichment should be revisited. 

The main contributions of this paper are:
\begin{itemize}
    \item A discussion of current issues in bridging qualitative and quantitative methods analysis and how modern AI could help.
    \item A model of the transitioning between a data and semantics perspective in the process of analyzing data-rich user \mbox{study observations.}
    \item A discussion of open issues and research opportunities in the interplay between AI, human analyst, and qualitative and quantitative methods.
\end{itemize}

\section{Related Work}
\label{sec:relatedwork}

This paper is embedded in the long-standing discussion of research methods for visualization. General background information on this topic is provided, for example, by Carpendale~\cite{Carpendale:2008:EvaluatingInformationVisualizations}. She pays special attention to the role and usefulness of qualitative methods. For example, grounded theory \cite{Corbin:2008:BasicsQualitativeResearch,Glaser:1992:BasicsGroundedTheory} or qualitative content analysis~\cite{Mayring:2014:QualitativeContentAnalysis} are often employed. Scenarios of evaluation in visualization research are summarized by Lam et al.~\cite{Lam:2012:EmpiricalStudiesInformation} and Isenberg et al.~\cite{Isenberg:2013:SystematicReviewPractice}. Kurzhals et al.~\cite{Kurzhals:2016:EyeTrackingEvaluation} cover more concrete examples of eye tracking evaluation, particularly for visual analytics. Design studies~\cite{Sedlmair:2012:DesignStudyMethodology} are typical examples where qualitative methods are often applied. Here, special attention should be paid to the issue of how rigor can be established~\cite{Meyer:2020:CriteriaRigorVisualization}, e.g., through abundance and transparency of data acquired through the design study process and reported later. Overall, there has been substantial progress in how visualization research is evaluated---with a plurality of methods. This paper builds on this basis and provides discussion points that could lead to rethinking some of the roles and connections between different evaluation methods.

In particular, I address the challenges and benefits of bridging qualitative and quantitative methods. The underlying general theme of mixed methods is covered, for example, in several books~\cite{Johnson:2007:DefinitionMixedMethods,Tashakkori:1998:MixedMethodology}. Typical approaches include triangulation, embedded design, explanatory design, and exploratory design (see Creswell and Plano Clark~\cite{Creswell:2007:DesigningConductingMixed}). I focus on the integration of techniques applied to the same common study design and on data integration there. An overview of approaches for data integration in mixed methods research is provided by Caracelli and Greene~\cite{Caracelli:1993:DataAnalysisStrategies}, data integration for convergent studies is discussed by Moseholm and Fetters~\cite{Moseholm:2017:ConceptualModelsGuide}, and data integration and consolidation by Vogl~\cite{Vogl:2019:IntegratingConsolidatingData}. This paper fits into the general scope and objectives of mixed methods, taking a new perspective on the tight interplay between quantitative and qualitative perspectives using AI. Koch et al.~\cite{Koch:2023:VisualizationPsychologyEyetracking} also discuss bridging between qualitative and quantitative research methods, for the special case of eye tracking evaluation. However, their work focuses more on the role of visualization psychology and does not include AI in combining research methods.

I take the ongoing advancements in AI as a basis for revisiting the connection between quantitative and qualitative methods. A very recent trend is to make qualitative research more efficient and less labor-intensive by using AI techniques. The focus is on using LLMs because they allow one to address the many different facets of text analysis needed in various qualitative methods. Roberts et al.~\cite{Roberts:2024:Artificialintelligencequalitative} provide a critical discussion of the potential of LLMs for qualitative research in general. As more concrete examples, there is recent work on supporting qualitative analysis via deductive coding with ChatGPT \cite{Xiao:2023:SupportingQualitativeAnalysis}, the use of ChatGPT for qualitative coding \cite{Zhang:2023:QualiGPTGPT}, LLM-based analysis of text from online forums~\cite{Rao:2024:QuaLLM}, or LLMs for thematic analysis \cite{Dai:2023:LLMloopLeveraging}, which is useful as part of qualitative analysis. Bano et al.~\cite{Bano:2024:AIHumanReasoning} assess LLMs compared to human judgment for qualitative research. LLMs may even be used as interview partners~\cite{Dengel:2023:QualitativeResearchMethods} for qualitative work, going beyond coding and analysis. The papers above relate to helping analyze qualitative study data but do not fundamentally change the research methodology itself. This is where this paper differs: using LLMs in particular, and AI in general, allows for rethinking the interplay between quantitative and qualitative perspectives for visualization research. Karjus~\cite{Karjus:2023:MachineAssistedMixed} provides a related perspective from the humanities and social sciences on AI for mixed methods. A recent workshop addressed LLMs for human-computer interaction research in general~\cite{AubinLeQuere:2024:LLMsasResearch}.

This paper is positioned in the larger context of AI support for scientific methods. 
Messeri and Crockett~\cite{Messeri:2024:ArtificialintelligenceIllusions} discuss AI for scientific research in general. They identify four visions of AI: ``AI as Oracle'' (supporting the study design stage of research), ``AI as Surrogate'' (for generating surrogate data to facilitate the data collection stage), ``AI as Quant'' (improving the data analysis stage, especially for large and complex data), and ``AI as Arbiter'' (for AI-supported peer review). This paper addresses the stage of data analysis, i.e., a specific case of ``AI as Quant.'' Messeri and Crockett discuss issues of AI in scientific work, including illusions of explanatory or exploratory depth, illusion of objectivity, and the overarching problem that the ``proliferation of AI tools in science risks introducing a phase of scientific enquiry in which we produce more but understand less.'' Such issues need to be considered for AI-supported research in general, and I will also discuss them for the special case of bridging quantitative and qualitative methods. 

The integration of interactive visualization, visual analytics, and AI and ML techniques is key to supporting quantitative and qualitative data analysis. The visualization and visual analytics community has been addressing the general topic of the connection to AI and ML extensively, with much ongoing work. Accordingly, there are many respective survey papers, covering visualization for ML~\cite{Sacha:2019:VIS4ML}, ML or AI for visualization~\cite{Wang:2022:SurveyML4VIS,Wu:2022:AI4VIS}, deep learning for scientific visualization~\cite{Wang:2023:DL4SciVis}, the integration of ML in visual analytics~\cite{Endert:2017:StateArtIntegrating}, and the role of generative AI in the context of visualization~\cite{Basole:2024:GenerativeAIVisualization,Schetinger:2023:DoomDeliciousnessChallenges}. In addition, there is work on AI in the context of natural language interfaces for the particular case of visualization~\cite{Shen:2023:TowardsNaturalLanguage}, and work on human-AI teaming and human-centered AI~\cite{Shneiderman:2022:HumanCenteredAI}. These general research trends will be instrumental in facilitating the next level of mixed methods research but will require adaptation to the special needs of empirical research data.

\section{Background and Motivating Examples}
\label{sec:backgroundmotivating}

This section provides some examples of own previous work that motivate a new perspective on bridging qualitative and quantitative research. This previous work was developed as part of a larger collaborative research center on ``Quantitative Methods for Visual Computing''\footnote{Transregional Collaborative Research Center SFB-TRR 161 (``Quantitative Methods for Visual Computing''), \url{https://www.sfbtrr161.de/}} \cite{Schreiber:2022:QuantitativeVisualComputing}. This center brings together researchers from all areas of visual computing: visualization, computer graphics, human-computer interaction, computer vision, and image processing. The disciplines have their own set of empirical research methods that span a wide range from very controlled quantification methods and benchmarking to qualitative and design-oriented work. There are traditional quantitative methods with NHST but also fine-grained measurements for psychophysical or physiological research~\cite{Scheer:2016:SteeringDemandsDiminish}, development of benchmarks and generative dataset~\cite{Schulz:2016:GenerativeDataModels}, detailed technical measurements~\cite{Mueller:2022:PowerOverwhelmingQuantifying}, coding for qualitative investigations~\cite{Angerbauer:2022:AccessibilityColorVision}, or case studies and application-oriented visualization design~\cite{Schaetzle:2019:VisualizingLinguisticChange}. Combining this range of expertise provides opportunities to advance research methods for visualization, as briefly illustrated in this section. It should be noted that the research center uses ``quantitative methods'' just as a starting point for a broader perspective on evaluation methods as well as replicability and reproducibility~\cite{Garkov:2022:ResearchDataCuration}.

\begin{figure}[t]
    \centering
    \includegraphics[width=1.0\linewidth]{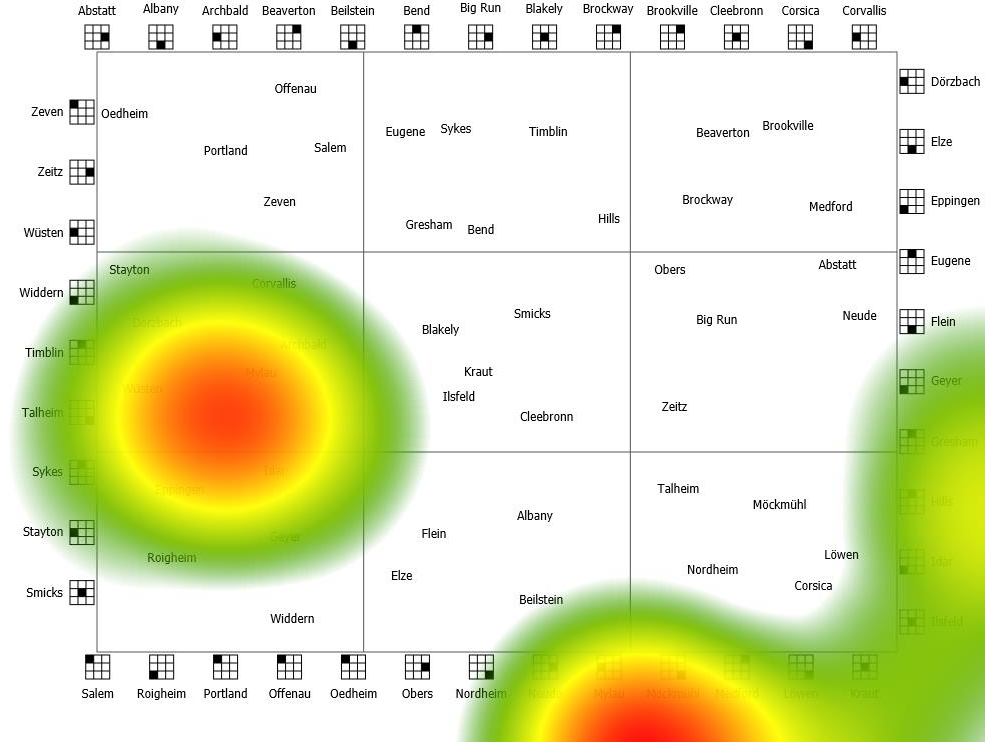}
    \caption{Heatmap of the distribution of attention (here, for a user study on visual search in maps).
    Image \copyright~2017 IEEE. Reprinted with permission from Netzel el al.~\cite{Netzel:2017:EvaluationVisualSearch}.}
    \label{fig:heatmapvisualsearch}
\end{figure}

I want to pick eye tracking as one example that inherently carries quantitative and qualitative aspects, which can be characterized as that the ``evaluation of eye tracking results in qualitative data, in the form of Heat Maps and Gaze Plots, as well as quantitative key figures through the calculation of metrics"~\cite{InnovationAcceptanceLab:2024:ETT}. There is literature covering the background on eye tracking in general~\cite{Duchowski:2017:EyeTrackingMethodology,Holmqvist:2011:EyeTracking} and the visualization of eye tracking data in particular~\cite{Blascheck:2017:VisualizationEyeTracking}. First of all, it should be noted that eye tracking provides substantial amounts of data, e.g., by measuring gaze at sampling frequencies typically between 60\,Hz and 2,000\,Hz. While statistical summarization or inference are typically used in controlled laboratory experiments for psychology research, other disciplines often reduce the analysis and reporting to just showing attention heatmaps or gaze replays. In this sense, quantitative data is interpreted and analyzed by a typical qualitative approach. 

\Cref{fig:heatmapvisualsearch} shows an example of a heatmap visualization of the distribution of visual attention for the case of a controlled experiment. This example demonstrates that the underlying stimulus is critical for understanding the attention heatmap. Here, a cue for supporting visual search was given at the bottom part of the image (therefore, we see an attention hotspot there) and the final search target was in the center-left of the image (with the other hotspot). Only with this relevant information about the stimulus and task can the attention visualization be adequately interpreted.
Another example is the use of eye tracking visualization in qualitative studies, where gaze visualization is integrated into coding systems like ChronoViz to facilitate manual coding of study data \cite{Weibel:2012:LetsLookCockpit}.

Another approach adopts a quantitative perspective by applying (sometimes several different) eye tracking metrics to gaze data~\cite{Holmqvist:2011:EyeTracking}, thus facilitating statistical inference, i.e., traditional quantitative analysis (see an example in the work by Netzel et al.~\cite{Netzel:2017:EvaluationVisualSearch}). However, these metrics often lack reliability in their interpretability; for example, they might be used to indicate cognitive load but with quite some associated ambiguity and inaccuracy.

\begin{figure}[t]
    \centering
    \includegraphics[width=1.\linewidth]{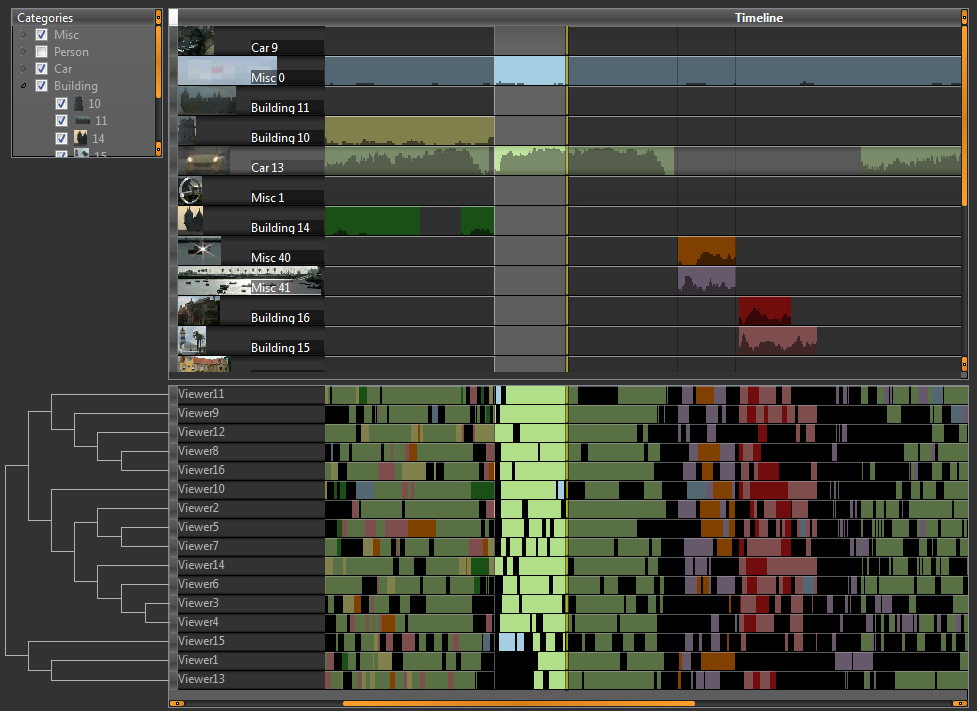}
    \caption{Eye tracking study data with enriched semantics: (top)~Annotation by AOIs provides meaning to glanced-at regions of the stimulus, along with quantitative data about their relative frequency over time. (bottom/center-right) Scarf plots with the same color coding as for the AOIs above show the distribution of attention for each individual participant over time. (bottom/left) Dendrogram of hierarchical clustering of scanpaths for the participants. Image \copyright~2014 ACM, reused from Kurzhals et al.~\cite{Kurzhals:2014:ISeeCube}.} 
    \label{fig:scarfclustering}
\end{figure}

Therefore, there are alternative methods that keep more of the qualitative and quantitative aspects of eye tracking data along a longer analysis pipeline. For the example of the \emph{ISeeCube} system~\cite{Kurzhals:2014:ISeeCube}, the process and data typically comprise the original gaze data, an assignment of gaze to areas of interest (AOIs) from the underlying image that carry semantic information (meaning of attended parts of the stimulus), frequencies of attendance to AOIs over time, scanpath representations of viewing behavior, and an analysis of typical scanpath behaviors as a basis for identifying reading strategies. \Cref{fig:scarfclustering} illustrates several of these representations. The key observation is that the assignment of AOIs (and their semantics) adds meaning to the gaze data. This is similar to other coding in qualitative research. However, we still have quantitative information about attended AOIs over time, but now in a more aggregated and semantically enriched form compared to the raw gaze data. In the subsequent step, we can apply quantitative techniques---here, hierarchical clustering based on distance metrics for scanpaths---to identify patterns of gaze and participant behavior.

\begin{figure*}
    \centering
    \includegraphics[width=0.8\linewidth]{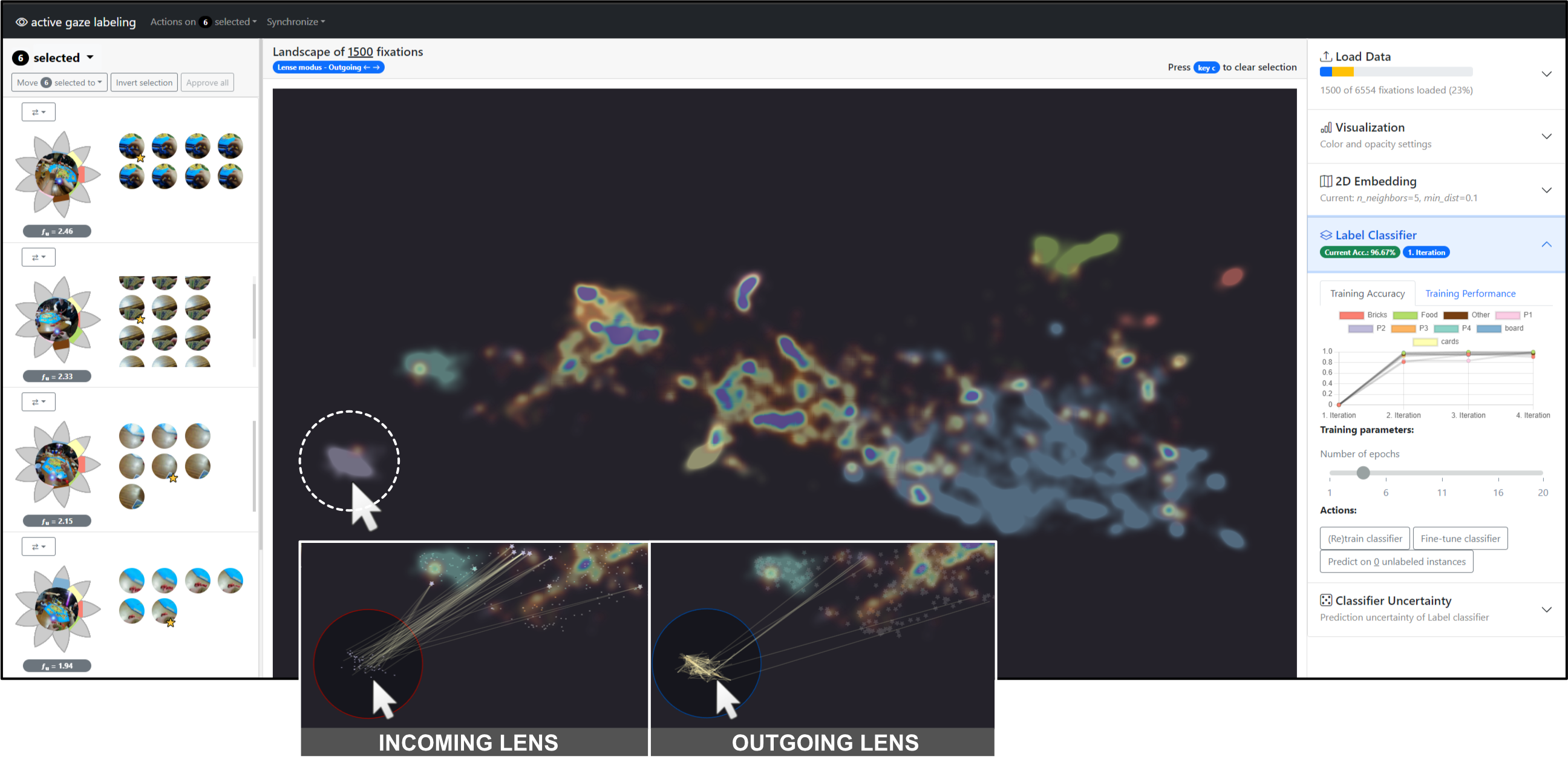}
    \caption{Interactive visualization system for trust building in AI-assisted labeling of video data associated with eye tracking recordings: (left)~glyph-based visualization of prediction uncertainty of the ML classifier, (center)~2D embedding of image thumbnails around gaze positions, along with zoomed-in insets showing the interactive exploration of fixation details for incoming and outgoing connections, (right)~panel with application settings and information.
    Image reprinted from \copyright\ 2024 Koch et al.~\cite{Koch:2024:ActiveGazeLabeling}, licensed under CC BY 4.0.}
    \label{fig:activegazelabeling}
\end{figure*}

A study on evaluating participants' strategies for reading metro maps~\cite{Netzel:2017:UserPerformanceReading} illustrates some of the interplay between qualitative and quantitative methods. This study includes statistical inference directly applied to eye tracking metrics. However, to get a deeper understanding of participants' reading strategies, we used qualitative exploration of scanpath visualizations in the pilot phase to arrive at a quantitative description in the form of univariate distance plots and respective aggregation by the bimodality coefficient for statistical inference. This example shows that there can be several steps of data transformation, informed by qualitative understanding, to eventually obtain a statistical result. This study also shows the large time effort involved in manual coding: For yet another perspective on participants' reading strategies, we had to develop a specialized annotation tool that relied on eye tracking visualization to support an annotator in disambiguation of gaze on metro lines (which can be very close to each other on the stimulus) for fixation labeling~\cite{Netzel:2016:InteractiveScanpathOriented}; the annotated fixations were  then used for another statistical analysis. Even with this tool, it took about 140 hours to annotate 39,404 fixations.

The metro map study took place in a controlled laboratory environment, relied on a few static stimuli only, had clear tasks, and limited task duration lengths. However, even such a controlled study already needed a combination of quantitative and qualitative approaches to capture the different facets of the study that come with several modalities (task performance measures, stimulus, and eye tracking recordings). Furthermore, the study required substantial manual annotation of eye tracking data. This effort involved in coding demonstrates the need for better machine support.

The sketched problems become much more pronounced once going to more complex scenarios, such as lesser controlled studies ``in-the-wild,'' experiments that include even more modalities of recorded information, long-term studies, or studies that are traditionally analyzed by qualitative methods. Jung et al.~\cite{Jung:2018:MethodologicalCaseStudy} stated in 2018: ``The advantages of eye-tracking have been overlooked by qualitative researchers, as most eye-tracking studies utilized quantitative approaches [...]. However, unlike eye-tracking, mobile eye-tracking can provide detailed contextual information, which is critical for qualitative research.''

Similarly, we have included eye tracking in studies not traditionally enriched by gaze information. One example concerns studies that include same-place/same-place collaboration of participants, such as pair programming~\cite{Kumar:2020:DemoEyeSACSystem}. Other examples are the analysis and annotation of user activity, here with long-term pervasive eye tracking with durations of several hours~\cite{Kurzhals:2020:VisualAnalyticsAnnotation}, or studies with mobile eye tracking in augmented reality~\cite{Oeney:2023:VisualGazeLabeling}. A number of ongoing studies use unconstrained setups, with several participants interacting, with several modalities being recorded (including eye tracking, participant-centric and external camera recordings, and audio), and for longer durations (about one hour per experiment).

For data analysis, we use a combination of approaches. Similar to previous work by others (see \Cref{sec:relatedwork}), we have had good experiences with AI-based text-to-speech and text summarization to transcribe and summarize audio recordings of participants' comments. A challenge is the integration of eye tracking and video data for unconstrained settings. Here, we have been working on integrating interactive visualization with computer vision techniques for annotating and analyzing video thumbnails around gaze locations. One example is our work on active gaze labeling~\cite{Koch:2024:ActiveGazeLabeling}, which facilitates uncertainty-aware visual representations to build trust in computer vision classifiers based on examples of annotated samples. \Cref{fig:activegazelabeling} shows a screenshot of the system. With this approach, we can substantially reduce the effort involved in traditional manual annotation, and still give enough control to the analyst for reliable and trustworthy results.

\begin{figure*}
    \centering
    \includegraphics[width=0.77\linewidth]{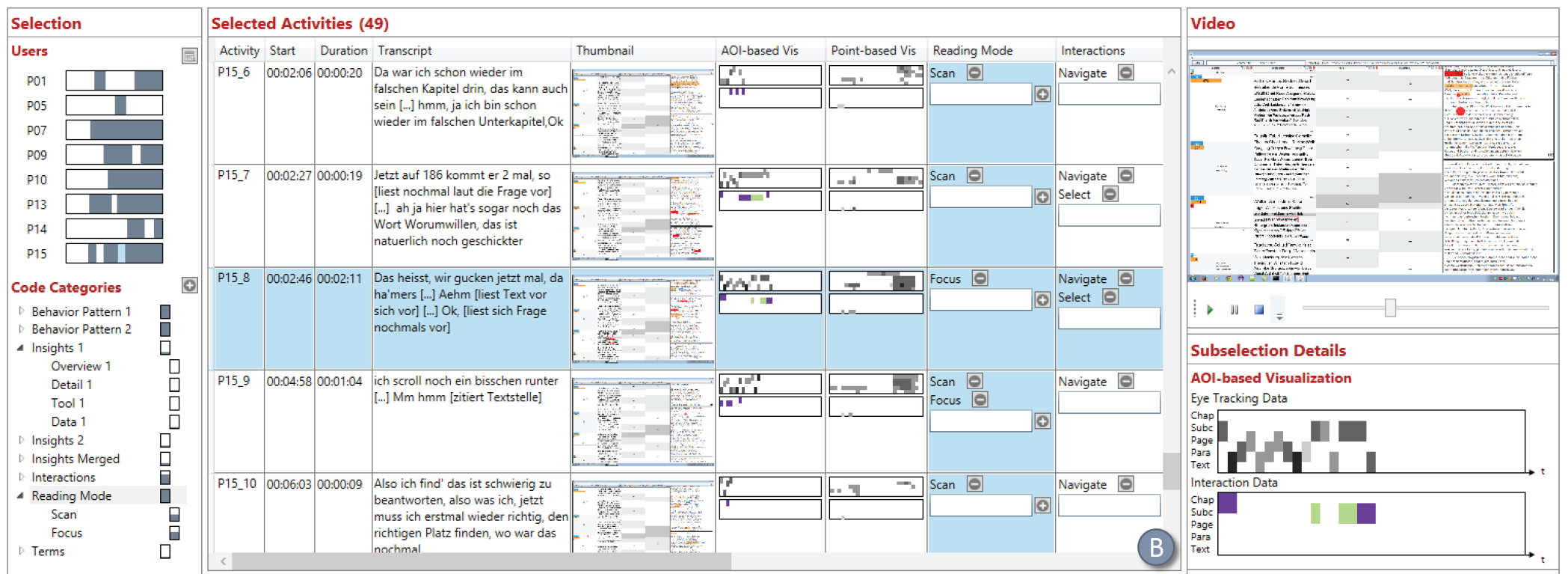}
    \caption{Visual interface for coding of data-rich user behavior, including transcribed text, several facets of eye tracking data (point-based data and AOIs), and interaction logging. Image \copyright~2016 IEEE. Reprinted with permission from Blascheck el al.~\cite{Blascheck:2016:VisualAnalysisCoding}.}
    \label{fig:richcoding}
\end{figure*}

In general, the underlying studies are evaluated with quantitative and qualitative approaches to obtain an in-depth understanding of the results. Coded data is often analyzed by statistical techniques but also aligned with qualitative summarizations for explanations. For some user studies, we used fully manual, traditional coding of multimodal observations. \Cref{fig:richcoding} shows the screenshot of a visual interface that supports coding such data-rich user behavior~\cite{Blascheck:2016:VisualAnalysisCoding}. The interface also facilitates higher-level coding, abstracting the original, low-level codes to higher-level ones as part of a second coding cycle, i.e., there can be a two-level hierarchy.

These examples show that there are many benefits from combining traditionally separate techniques from qualitative and quantitative evaluation and analysis, leading to a deeper understanding of empirical evaluation, in particular, for less constrained setups like ``in-the-wild'' studies. However, a big challenge is that substantial effort is involved in the manual stages of annotation, coding, or other interventions by analysts. Therefore, many approaches are hard to implement in daily practices. Our encouraging experiences with the integration of computer support motivate the further use of AI in these processes.

\section{Role of AI}
\label{sec:AI}

As discussed in \Cref{sec:relatedwork}, LLMs are already capable of improving traditional qualitative work (like protocol analysis, grounded theory, or coding) that is heavily based on text analysis. However, advances in AI do not stop with language models. In particular, there has been much progress in computer vision, which is equally important for visualization research methods because many of the study data exhibit images or related data, e.g., in the form of recordings by exterior cameras or egocentric ones carried by participants. Even more important is AI research on multimodal data, such as addressing language and image models jointly. 
Therefore, it is becoming increasingly possible to use AI techniques for data-rich observations from user studies.

For example, progress in multimodal LLMs can play an important role in working with data-rich and diverse recordings from user studies, including audio, images, or video. Surveys of multimodal LLMs are provided by Yin et al.~\cite{Yin:2023:SurveyMultimodalLarge} and Zhang et al.~\cite{Zhang:2024:MMLLMs}. Multimodal LLMs benefit from the progress in LLMs in general (see the survey by Zhao et al.~\cite{Zhao:2023:SurveyLargeLanguage}) but also from advances in AI for the other modalities---often from the area of computer vision \cite{Awais:2023:FoundationalModelsDefining,Madan:2024:FoundationModelsVideo}.

With modern AI, we have promising means toward closing the gap between---sometimes semantically poor---input data and semantically rich interpretations and insight building from user studies. The flexibility and generalizability of foundation models, benefits of in-context learning, chain-of-thought, or instruction tuning, and approaches to few-shot or zero-shot learning pave the way for supporting data analysis from studies in uncontrolled environments. Even if AI alone might not be able to fully provide semantic information, there are opportunities for efficiently including semantics by combining human expertise and input with AI. However, many open issues still need to be addressed before AI can take a central role in helping analyze qualitative data. One issue is that the quality of AI might depend substantially on the context and type of data. Improved generalizability and adaptability are primary goals of AI research in general. Therefore, it is foreseeable that AI techniques can handle more and more areas of user study data. However, there might still be the need to contribute some particular aspects from the perspective of visualization research. For example, the use case of eye tracking data might need the extension of existing or future multimodal LLMs to include this specific modality.

Another issue is the extent of reliability, controllability, and explainability of AI. In particular, hallucinations of the AI should be avoided for trustworthy data analysis. Of special interest could be extensions that would link AI results to evidence in the input data, thus providing a traceable path from raw data to AI-generated insights. There is also quite some research in these directions in the AI community. Therefore, a positive impact on visualization evaluation methods can be foreseen. Furthermore, I expect that there will be ongoing research in the visualization community and outside, on integrating humans in AI processes for trustworthy, explainable, and controllable results, often in the context of human-AI teaming.

Finally, both symbolic and subsymbolic approaches are heavily researched in AI, which could also help visualization evaluation because of the need for explicit knowledge modeling and representation as well as rather computational, data-centered techniques. In particular, neural-symbolic AI (see, e.g., the survey by Yu et al.~\cite{Yu:2023:SurveyNeuralSymbolic}) could play a role in, at least some parts of, the data analysis for empirical visualization research. It should also be pointed out that AI support should not stop at making coding more efficient. In fact, there is a great opportunity that AI can help improve other sensemaking processes, albeit with the need for special attention to the limits of AI and the potential illusions of understanding in scientific practice~\cite{Messeri:2024:ArtificialintelligenceIllusions}.

\begin{figure*}
    \centering
    \includegraphics[width=0.59\linewidth]{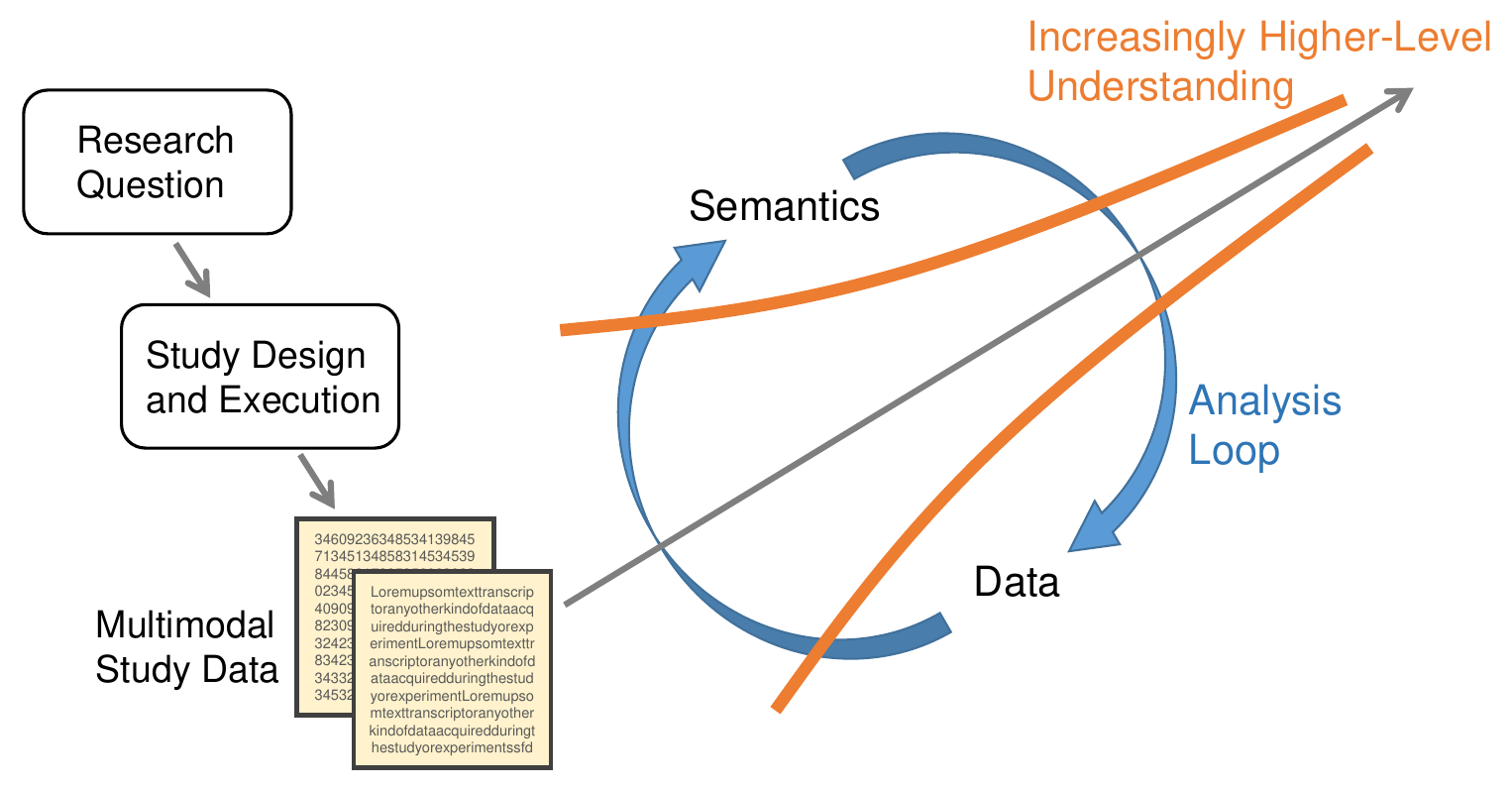}
    \caption{Schematic process of research question, study design and execution, and iterative analysis of (possibly multimodal) study data. The key part is the analysis loop that keeps on transforming and enriching data with additional semantics to derive new data representations.  Through the process, information is obtained at higher and higher levels of understanding. The analysis loop may consist of AI-based processing, user intervention, or a combination thereof.}
    \label{fig:process}
\end{figure*}

\section{Process Model}
\label{sec:processmodel}

Let us take the general motivation for bridging qualitative and quantitative methods (from \Cref{sec:backgroundmotivating}) and combine it with current and expected future improvements in AI (see \Cref{sec:AI}). Then, the process illustrated in \cref{fig:process} becomes feasible. It starts with a research question (as usual) and the design and execution of the study. The latter might be influenced by the available capabilities of subsequent data analysis. For example, if particular sets of sensor data become manageable by advanced AI analysis, it might be attractive to consider acquiring such data in the study. 
The user study should make use of data-rich observations where possible. Therefore, the default is that there are multiple, multimodal observations, such as audio or video recordings of the study, transcripts of think-aloud protocols, logged mouse and keyboard interaction, eye tracking data, or recordings of other physiological sensor data. 

The key point of the process is the analysis loop that iteratively enriches data with additional semantics to arrive at new data, ideally at a higher level of understanding. Semantic enrichment may be done through coding, topic analysis, identification of causal relationships, etc. 
Koch et al.~\cite{Koch:2023:VisualizationPsychologyEyetracking} describe such a step of qualtizing as a ``semantic transformation of the original quantitative data.'' We may also use transformations that observe the relevant semantics of the input representation to arrive at a different, often more condensed representation. A typical example is the transcription of an audio recording, which transforms the audio signal into a much compressed written form. Condensed representations are obviously useful when later processed by human analysts, but they might also help by allowing the execution of AI techniques designed for the target representation. The analysis loop allows for several transitions between the representations of observations. 

\paragraph{Relationship between data and semantics} Although \emph{data} and \emph{semantics} are the two key elements of the analysis loop in \cref{fig:process}, they do not constitute a dichotomy but rather work hand in hand.  We need schemas (concepts) to handle and process data; the semantics assigned to data links to the underlying concept. There is a related discussion in the context of Iceberg Sensemaking~\cite{Berret:2022:IcebergSensemakingProcess}, highlighting the role of schemas in (visualization and data) sensemaking. Another related aspect is the philosophical perspective of action theory addressing how we can recognize intent in action~\cite{Feige:2023:PhilosophyActionRelationship}. I want to avoid fundamental philosophical discussions, as often covered in social sciences or epistemology; more references to these backgrounds can be found in the above paper~\cite{Berret:2022:IcebergSensemakingProcess}. Instead, I take a pragmatic approach by focusing on practical implications and opportunities for improving research methods in the era of massively advancing AI. 

Already the ``raw,'' often multimodal study data comes with semantics. For example, eye tracking data might consist of floating-point numbers with the semantics of $x$ and $y$ coordinates of gaze points over continuous time $t$. By assigning AOIs, we can enrich such time-dependent data with the semantics of the underlying stimulus. At the same time, the data representation is changed: from a time series of $x$ and $y$ coordinates to a time series with temporally equidistant samples of AOI labels. The first iteration of semantic enrichment is present in many examples of qualitative research, in particular, by assigning codes (or AOIs, as in the above example). However, we might want to continue by taking the derived data (here, the time series of AOI labels) and perform coding of user activity, leading to a sequence of time spans annotated by activity. Therefore, the resulting data has a different representation and semantics (a sequence of time spans of differing lengths and associated with activity labels) than the previous time series of AOIs. During the iterative process, we typically perform some aggregation or summarization of data (in the example, AOI labels sampled at, e.g., 120\,Hz, are aggregated to a sequence of time spans of several seconds of user activities) and/or merging of multiple dataset (in the example, implicitly merging the image stimulus information with the gaze data by defining AOIs). In this way, we can arrive at a higher level or more abstracted understanding of the study data, as indicated by the narrowing orange curves in \cref{fig:process}.

\paragraph{Types of representation and blurring qualitative and quantitative perspectives}
It should also be mentioned that, while sequence or time-oriented data is quite common for user studies, data types are not restricted to these. They might also include hierarchical information of aggregation and clustering results, or graphs representing causal or other relationships. Therefore, some of the data might be suitable for immediate application of traditional statistical analysis (e.g., for comparing frequency/counts of occurrences), whereas other data might need other types of quantitative analysis. In particular, later stages of the analysis loop might come with a large variety of potential forms of representation, including graphical models, sequences, point sets in high-dimensional space, text, or images.

Depending on the type of representation, the boundaries between quantitative and qualitative perspectives will become blurred. For example, short texts (e.g., participant comments or microblogging documents) might be considered text collections for traditional qualitative analysis. If the same input is represented as a collection of high-dimensional points via text embedding by natural language processing, we can still perform similar tasks as with qualitative manual analysis (for example, clustering into groups of similar texts), but we will also have metrics to measure similarity between texts. It is not obvious whether a text embedding should be seen as a representation for qualitative or quantitative analysis. I think that the boundaries become less relevant once we transition to more complex forms of representation. In general, AI techniques often come with metric information, for example, for latent space representations, or as for the above example of embedding spaces for text. Therefore, quantitative aspects are not only related to counting occurrences (such as counting for coded user study data) but also to measurements of distance, similarity, relevance, weight, etc.

\paragraph{Role of humans and AI}
The analysis loop in \cref{fig:process} may be performed manually by a human analyst (e.g., the traditional approach of coding), fully automatically by AI, or by combining partly human and partly AI-based semantic transformation or enrichment. The process of \cref{fig:process} leaves open how the work is distributed between humans and AI. In particular, the distribution may vary for each iteration of the analysis loop. Currently, low-level transformations or semantic enrichment (first iterations) might be more suited for completely AI-based processing (e.g., speech-to-text transcriptions of audio recordings) than later phases, where more interpretation by the analyst might be required. Such balancing and adaption of the roles of humans and AI are also in line with the recommendations by Messeri and Crockett~\cite{Messeri:2024:ArtificialintelligenceIllusions}, who see the use of AI especially for routine tasks and those within one's area of expertise. In general, there should be mechanisms for trustworthy AI processing and controllability by the analyst. For example, there should be methods for data drill-down to confirm findings with the input data, as well as traceability and provenance of AI results. Furthermore, AI systems---similar to humans---might include subjective interpretations. Therefore, checks not only by humans but also by different AI systems (for AI agreement) might be useful.

Human-AI teaming could be a typical approach for the higher levels of processing and abstraction, allowing the analyst to include their tacit knowledge and turn it into explicit knowledge (e.g., via assigning codes). With the advances in AI, there will be an ongoing adjustment of the roles of humans and AI, with a general trend toward increasing automation and higher and higher levels of semantics included in AI models.

\paragraph{Number of iterations of the analysis loop}
There is no general rule on how often the analysis loop should be iterated. In fact, semantically enriched data may be used at any stage or iteration of the process. It could already be in the first cycle, e.g., by performing statistical analysis of the frequency of occurrences of some coded low-level behavior. It could also be higher-level, more abstracted data from later iterations, e.g., frequencies of higher-level codes or hierarchically clustered participant behavior according to some comparison metric. Even for the same user study, extracting data and semantics from different analysis cycles might be useful to get different perspectives on the study. For example, it might be instructive to understand the distribution of low-level task execution by participants, and, at the same time, it might be important to understand their high-level strategies. It should also be noted that, when taking semantically enriched data out of an analysis cycle, it might still undergo a quantizing process~\cite{Sandelowski:2009:Quantitizing} to make it useful for statistical testing, e.g., by counting occurrences of codes~\cite{Koch:2023:VisualizationPsychologyEyetracking}. 
The iterative analysis loop also comes with the benefit that it allows including plausibility checks and grounding to analysts' experience and expert knowledge. I do not expect that the complete process is fully automatic and based on AI. Instead, there are assessments by human judgment involved, thus addressing issues of controllability and trust and reducing the risk of AI illusions and hallucinations. Furthermore, each cycle should rely on an adequate choice of AI techniques and an appropriate balance between human and AI roles, which will typically differ from cycle to cycle.

\vspace{1.5ex}

Overall, the main message of the process model of \cref{fig:process} is to use data-rich, possible multimodal observational data and iteratively enrich it with additional semantic information.
I find this process model and the focus on data/semantics useful for framing and guiding the development of methods for bridging qualitative and quantitative approaches. However, the proposed process model should be understood just as one proposition, and it is up to discussion how other frameworks could provide different benefits and ways to integrate AI into research methods.

\section{Discussion and Outlook}

This section discusses the roles of qualitative and quantitative methods in visualization research, open issues, and opportunities.

\paragraph{Importance of bridging qualitative and quantitative methods for visualization}
While research methodology in general, and mixed methods in particular, are important for many disciplines, they are especially relevant for visualization research. One reason is the large design space typically associated with visualization research. It is pretty uncommon that research questions can be broken down into simple comparisons in analogy to treatment vs.\ no treatment in medical research. Instead, many parameters could influence the visualization or the interaction with the visualization. For such problems, traditional, well-controlled quantitative experiments are too restrictive. This is particularly true for studies ``in-the-wild'' and others with good ecological validity. Therefore, there is a need to include a more flexible and semantically enriched analysis of observations. At the same time, quantitative assessments can shed additional light on how participants work with complex visualization, leading to a more in-depth understanding and increased validity of studies. Another reason is the dual role of user and machine and their interplay. We want to understand both the human user and the technical realization of the visualization. In contrast, much of psychology is concerned with understanding the human (mind). Similarly, ethnography focuses on (groups of) humans. At the other end of the spectrum, typical engineering research addresses the evaluation of technology. In contrast, visualization research is often concerned with understanding humans, technology, and their interplay. This bridging of the different realms should also be reflected by the employed research methods.

\paragraph{Beyond just making existing qualitative research methods faster} As discussed in \Cref{sec:relatedwork}, many of the current applications of AI aim to make existing qualitative research methods faster, which is an important step ahead. However, it would be a missed opportunity to stop there. Once we have easy access to more quantitative coding and rich sets of observations, we may explore analysis and inference techniques beyond statistical testing or qualitative summarization. A promising direction is considering causality identification~\cite{Guo:2022:SurveyVisualAnalysis}. For example, causality and sequence analysis, which has been applied to sequences of electronic health records or comments in online media \cite{Jin:2021:VisualCausalityAnalysis}, may be extensible to sequential coding information from (multimodal) user study data. A further extension could even include causality identification for multiple outcomes~\cite{Fan:2025:VisualAnalysisMultioutcome}.
Such semi-automated techniques could help uncover much more complex structures observed in empirical evaluation---in line with the complexity of the problem setting of large design spaces discussed above. Another direction could be to include a more quantitative flavor in areas where qualitative approaches are mostly used. For example, design studies~\cite{Sedlmair:2012:DesignStudyMethodology} are hard to align with traditional goals of reproducibility, but they can be supported by abundant and transparent documentation~\cite{Meyer:2020:CriteriaRigorVisualization} that could benefit from automatic or semi-automatic analysis of textual or other documents. In particular, thick descriptions could be included and analyzed to arrive at a comprehensive understanding of design studies.

\paragraph{Evaluation in line with research and contribution type}
Visualization is particularly broad in its types of research and contributions~\cite{Lee:2019:BroadeningIntellectualDiversity}. This breadth has to be reflected by choosing appropriate evaluation methods. Therefore, it is advantageous if the available evaluation approaches and analysis techniques are extended as well, thus providing more choices in finding a good fit. The flexible process model of \Cref{sec:processmodel}, with its support for accessing quantitative and qualitative data, could be a starting point for a flexible and adaptable evaluation methodology. However, more research and best practices for underlying study design and analysis are still~needed.

\paragraph{Not restricted to evaluation alone}
AI techniques and quantification aspects might not only support evaluation but could pave a path to including findings on-the-fly in adaptive interactive visualization systems (see Chiossi et al.~\cite{Chiossi:2022:Adaptingvisualizationsinterfaces}). Once we have some machine-based computational or algorithmic processing of evaluation data, it is only a small step to access similar data during user interaction and apply the analysis in real time to let the system adapt to the user. 

\paragraph{Ethics, data protection, privacy, and open science}
The rich information that comes with multimodal and data-intensive observations not only opens up opportunities to better understand user studies but is also associated with additional issues of research ethics and data protection. Since we learn more about participants, there is the intrinsic danger of revealing personal information that could not be extracted previously. Therefore, adapted and improved methods for (pseudo) anonymization and privacy preservation need to be developed, along with best practices for obtaining appropriate informed consent and compliance with legal frameworks. These questions become especially relevant in light of the extended need for open science. However, with the proposed process model, there are also new opportunities. For example, ``raw'' data, like audio and video recordings of participants, tends to be problematic for publication, but more condensed, higher-level descriptions from the analysis loop might come with data that carries no personal information anymore and, thus, could be publishable. Furthermore, with generative AI, there are additional possibilities of anonymizing contents before public release (see, e.g., Kurzhals~\cite{Kurzhals:2024:AnonymizingEyeTracking}).

\paragraph{Evaluating evaluation}
When we develop new processes and techniques for evaluation, there is the question of the reliability of their results. Therefore, we need evaluations of these evaluation approaches. One way is to explore new methods for examples of studies that still include traditional, well-established evaluations for comparison. Furthermore, new benchmark datasets and tasks could foster the development of evaluation methods.

\paragraph{Call for action for the visualization community}
The above considerations indicate the expected benefits of bridging qualitative and quantitative methods for visualization research, which should alone be an incentive for the community to advance respective methods. Beyond this, the visualization research community is well positioned to work on these challenges because we have expertise in traditional qualitative and quantitative methods, human-centered computing, design, and much technology affinity and expertise. For example, there is already much work on combining human expertise with AI in various data analysis problems (see the survey papers on AI and ML in the context of visualization in \Cref{{sec:relatedwork}}). Following the idea of Vis4Vis \cite{Weiskopf:2020:Vis4Vis}, we are in an excellent position to address our own evaluation problems with new interactive visual analysis solutions, i.e., being our own domain experts. The call for action for the community is to work on improved research methods, best practices for study design, and corresponding ways of reporting results.

\section{Conclusion}

The main message of this paper is that, with the current advancements in AI, it is the right time to revisit and rethink the roles, characteristics, and interplay of qualitative and quantitative methods in visualization research. By including AI in bridging different approaches, we have the great opportunity to take evaluation methodology to the next level. There are many associated research problems that could be of great interest for visualization research on combining AI with visual analysis.

\acknowledgments{
This work was supported by the Deutsche Forschungsgemeinschaft (DFG, German Research Foundation)---Project ID 251654672---TRR 161.
Special thanks to Kuno Kurzhals and Maurice Koch for the many discussions on eye tracking, research methodology, and visualization. 
I also thank the reviewers of the paper for their very fruitful comments that helped improve the manuscript.}

\bibliographystyle{abbrv-doi-hyperref-narrow}

\bibliography{beliv2024}
\end{document}